\documentclass[aps,prl,preprint,superscriptaddress]{revtex4}
\usepackage{graphicx}
\begin{document}

%\preprint{}

\title{Graph equivalence and characterization via a continuous evolution of a physical analog}

\author{Vladimir Gudkov }
\email[]{gudkov@sc.edu}
%\thanks{}
\affiliation{Department of Physics and Astronomy
\\ University of South Carolina \\
Columbia, SC 29208 }
\author{Shmuel Nussinov}
\email[]{nussinov@ccsg.tau.ac.il}
 \affiliation{Tel-Aviv University \\
 School of Physics and Astronomy \\ 
Tel-Aviv, Israel \\
 and  \\ 
 Department of Physics and Astronomy\\
  University of South Carolina 
Columbia, SC 29208 }


\date{\today}

\begin{abstract}
A general novel approach mapping discrete, combinatorial, graph-theoretic problems onto ``physical'' models - namely $n$ simplexes in $n-1$ dimensions - is applied to the graph equivalence problem.  It is shown to solve this long standing problem in polynomial, short, time. 
\end{abstract}

%\keywords{}


\maketitle



\section{Introduction}

A graph G consists of $n$ vertices $V_i$ connected by edges $E_{ij}$. It is described by a connectivity
matrix $C$ with:
$$C_{ij}= C_{ji} = 0, 1  \hskip0.5cm  ( for \; (dis)connected \; V_i \; and \; V_j \;  i \neq  j =1, \cdots
, n)$$ 
\begin{equation}
C_{ii}= 0
\end{equation}
Vertex relabelling $i \rightarrow p(i)$ leaves G invariant but changes C according to
\begin{equation}
\label{eq:c}
C\longrightarrow  C' = P^T  C P
\end{equation}
with $P$ an orthogonal matrix with only one non-zero element in each row $i$ and column $j = p(i)$,
which represents the above permutation
\begin{equation}
\label{eq:p}
P = \delta _{(j , p(i))}
\end{equation}
The graph equivalence problem is the following: ``Given $C$ and $C'$, how can we decide, in time which
is polynomial in $n$, if both correspond to the same topological graph $G$ or to different graphs?,
or stated differently, does a permutation matrix $P$ for which Eq.(\ref{eq:c}) holds exist, and what is
this $P$ matrix?''

Exhaustive testing of all $n!$ permutation is impractical even for moderate $n$. A more systematic
search of $P$ performs just those transpositions which enhance an ``overlap'' function say
\begin{equation}
{\sl tr} C^{T} C' = \sum_{ij} C_{ij} C'_{ij}
\end{equation}
However the changes in $C$ (and in ${\sl tr} C^T C'$) due to any permutation is finite. There is no algorithm
for systematically enhancing ${\sl tr} C^T C'$, as subsequent transpositions may undo the improvement due to
previous permutations.

Our basic suggestion is: Instead of using discrete, large changes of say just two elements in a
transposition $(i \leftrightarrow j)$, we modify, in each step, all elements by small amounts.

Such ``continuous'' changes seem impossible: in the strict formal approach there are no ``continuous
permutations''.

\section{The Dynamical Model for Simplex Distortion}

We use a symmetric $n$ simplex (in $n-1$ dimensions) to represent our graph. The ``abstract'' vertices
$V_i $ of $G$ (or $C$) are mapped into the geometrical vertices $\vec r_i \ , \ \ \   i  =1, \cdots , n$
of the simplex. The symmetric configuration with all 
$|\vec r_i - \vec r_j | \ \ \  i \neq  j =1, \cdots , n$, equal, is the starting point of our algorithms.

The motion generated by the dynamics, was designed to distort the simplex by shifting its vertices from
the symmetric initial positions. The distorted simplex then reveals characteristic features of the
graph $G$ \cite{gold}.

The original aim of the distortion algorithm was to find groups of vertices in $G$ with higher than
average mutual connectivity, and asses the distances between the various clusters in the graph.

To this end attractive (repulsive) interactions were introduced between fictious point objects at 
$\vec r_i$ and  $\vec r_j$ when the corresponding vertices $V_i$ and $V_j$ are connected (or
disconnected) in $G$. We use first order ``Aristotelian'' dynamics:
\begin{equation}
\label{eq:for}
\mu \frac{d\vec r_i(t)}{dt}=\vec F_i (\vec r_i (t)),
\end{equation}
with forces $\vec F_i$ which derive from potentials:
\begin{equation}
\vec F_i = -\vec \nabla _{(\vec r_i)} \{U \left[ \vec r_i, \cdots, \vec r_n  \right] \},
\end{equation}
\begin{equation}
\label{eq:pot}
U = \sum_{i > j} U_a (| \vec r_i - \vec r_j |)C_{ij}+ \sum_{i > j} U_r (|\vec r_i - \vec r_j|)(1-C_{ij}),
\end{equation}
$U_a (r)$ ($U_r (r)$) are attractive (repulsive) pair-wise potentials.

By a proper tuning of the latter- which can even be modified as a function of ``time'' - we can
physically cluster at separate locations groups of points representing strongly (internally) connected
clusters in the graph $G$.

To avoid collapse towards the origin (or a ``run-away'' to infinity) if $U_a$ (or $U_r $) dominates, we
force $\vec r_i (t)$ to stay, at all times, on the unit sphere:
\begin{equation}
|\vec r_i (t)| = 1  \hskip1cm  all \hskip0.5cm t > 0.
\end{equation}
The graph characterization (G.C.) and graph equivalence (G.E.P.) problems are very closely connected.
If we could find (in polynomial number of steps!) a set of real numbers 
$\rho _1 , \rho _2, \cdots ,\rho _m $ that would completely characterize a graph $G$ then the G.E.P is
readily solved. All we need to do
is to compute for $C$ ($C'$) these numbers $\{\rho _k \}$ ($\{{\rho }'_k \}$), order the $\rho _k $ and
${\rho }'_k $ sets separately and compare them.

The set of eigenvalues $(\lambda _1, \cdots , \lambda _n )$ of the connectivity matrix are certainly
invariant under relabelling. While this set encodes a rich body of information of graph theoretic
interest, it fails to completely characterize graphs\cite{eigen}.

An alternative and natural simple variable helping characterize  connectivity matrices is the mutual entropy (see, for example \cite{prob}). 
Suppose the connectivity matrix $C$ has been normalized so that
\begin{equation}
\label{norm}
 \sum^n_{i,j=1}C_{ij}=1. 
\end{equation}
$P_i=\sum_j^nC_{ij}$ could then be considered as the probability that $V_i$ and $V_j$ are connected. The corresponding entropy 
\begin{equation}
\label{ent}
H(row) =- \sum^n_{j=1}P_{i}\log{P_{i}}, 
\end{equation}
could be considered as a measure of the uncertainty of the rows connection for the given network.  The amount of uncertainty for the connection of the column nodes given that the row nodes are connected is
\begin{equation}
\label{entc}
H(column|row) =- \sum^n_{i,j}C_{ij}\log{C_{ij}}-H(row).  
\end{equation}
As a result the amount of {\em mutual information} gained via the given connectivity of the network is
\begin{equation}
\label{inf}
I(C) =H(row)+H(column)-H(column|row),=\sum^n_{i,j}C_{ij}\log{(C_{ij}/P_iP_j)}. 
\end{equation}
where
\begin{equation}
\label{enf}
H(column|row)= - \sum^n_{i,j}C_{ij}\log{(C_{ij})}. 
\end{equation}
Due to the double summation and the symmetry of the connectivity matrix  $I(C)$ does not depend on the vertex relabelling and is a permutation invariant measure for the  connectivity matrix. 

Calculations of the mutual entropy for two connectivity matrices provides an  easy way to distinguish between these corresponding different graphs. If, however, the entropies are the same, the more detailed approach below is used. Amusingly we found that the entropy is already sufficient to distinguish between the lowest cospectral graphs (see, for example \cite{cosp} and references therein). 

The distances between the various vertices 
\begin{equation}
r_{ij}(t) \equiv |\vec r_i (t) - \vec r_j (t)|
\end{equation}
vary in our original algorithm as a function of time away from the original common value:
\begin{equation}
r_{ij}(0) = |\vec r_i (0) - \vec r_j (0)| = a \  \  \  \  \  \  \  all \ i \neq  j =1, \cdots , n
\end{equation}
Also in identical simulations of the dynamical evolution, the sets of relative distances computed for
$C$ and $C'$, should be the same if $C$ and $C'$ are equivalent:\\
\begin{equation}
\label{eq:rij}
\{r_{ij}(t)\} = \{r'_{ij}(t)\}  
\end{equation}
{\em One} permutation of $n$ elements (namely that which brings via Eqs.(\ref{eq:c}) and (\ref{eq:p}) $C$ into $C'$) should
yield:
\begin{equation}
\label{eq:dist}
|\vec r_{p(i)}(t) - \vec r_{p(j)}(t)| = |\vec{r'}_i (t) - \vec{r'}_j (t)|
\end{equation}\\
It is straightforward to verify (\ref{eq:rij}) and then using (\ref{eq:dist}) recover the permutation $i \rightarrow p(i)$.

In essence the idea of the present algorithm is to use the distortion of the simplex $S(0) \rightarrow S(t)$ 
\{i.e. $\vec r_i (0) \rightarrow \vec r_i (t)$\} generated via the dynamics of (repulsion) attraction
between (dis)connected vertices in $G$ to bring out an ``intrinsic shape'' of the graph.

Initially all vertices were at equal distances\cite{coor}. All the information pertaining to the graph was encoded
in the interactions of Eq.(\ref{eq:pot}).

After enough evolution steps, each vertex moves appreciably away, namely by\\
\begin{equation}
|\vec r_i (t) - \vec r_i (0)|\approx a/2
\end{equation}
from its initial position. The information on the specific graph $G$ reflects in the geometrical shape of
$S$,  i.e. the set of distances,
\begin{equation}
\label{eq:setd}
|\vec r_i (t) - \vec r_j (t)|  \hskip1cm  i \neq  j =1, \cdots , n.
\end{equation}
Vertices which are near in a graph theoretic sense, namely for which there are many, short, connecting
paths in the graph move closer together. (A short path consists of a small \# of consecutive edges which
starts at $V_i$ say and terminates at $V_j$). Like wise vertices which are far in a graph theoretic
sense i.e. have fewer and longer connecting paths will tend to move further away.

In our earlier work\cite{gjn} we sought to identify ``clusters in the graphs'' namely have the points
corresponding to a subset $\{C_i \}$ of vertices in the graph which have relatively strong mutual,
internal, connectivity, collapse to a single point.

For the present purpose we need (and should!) not pursue the evolution that far, as by then the graph
simplifies and some of the inter-cluster details are lost. Rather we need to stop ``Half-Way'': after
Eq.(\ref{eq:setd}) holds and yet no cluster has completely collapsed.

Note that in $n-1$ dimensions all the $n(n-1)/2$ distances $|\vec r_i (t) - \vec r_j (t)|$ are independent,
apart from triangular inequalities of the form
\begin{equation}
|\vec r_i (t) - \vec r_j (t)| \leq |\vec r_i (t) - \vec r_k (t)| +|\vec r_k (t) - \vec r_i (t)|. 
\end{equation}
Jointly these distances specify the geometric shape of $S$.

The mapping of the $n(n-1)/2$ bits of information: $C_{ij}=0 \ or \ 1$, via our dynamic evolution, into
the set of $n(n-1)/2$ distances, is highly non-linear.
The fact that we have $n(n-1)/2$ distances (rather than just $n$ eigenvalues) makes the former more
likely to specify the graphs.

Further we note that the time $t$ when the comparisons are made and the attractive and repulsive
interactions in Eq.(\ref{eq:pot}) above are free parameters and functions. Hence we can repeat the above graph
comparisons for many values and/or many functions $U_r (\rho )$, $U_a (\rho )$, making the significance
of a successful match extremely high.

If many of the $r_{ij} (t)$ $\{$and $r'_{ij} (t)\}$ are degenerate our ability to resolve graphs will be
diminished. However such degeneracies must stem from some symmetries in the graphs and corresponding
connectivity matrices. Once these symmetries are identified, the number of independent $C_{ij}$ (or
$C'_{ij}$) and the task of comparing them will be accordingly reduced.

We apply the above approach  below, demonstrating its power and
versatility.

\section{The convergence and complexity of the discrete modelings of the dynamical evolutions}

We follow the dynamics of the vertex shifts in Eq.(\ref{eq:for})  by discretizing the first
order equations:
\begin{equation}
\label{eq:discr}
\vec r_i (t + \delta ) = \vec r_i (t) + \frac {\delta }{\mu } \vec F_i (\vec r_e (t))
\end{equation}
with $\delta $ a small time increment.

Since $\vec r_i$, $\vec F_i$, are $n-1$  dimensional vectors Eq.(\ref{eq:discr})  represents
O($n^2$) equations for the relevant components.
Each force component $F_{i\alpha }$ is a sum of $v_i$ force
components with $v_i$ the valency of the vertex $V_i$ i.e. the \# of vertices connected to it. Hence
each step in (26) involves $n^2 v/2$ calculations with
\begin{equation}
v = \sum_{i =1} v_i /n
\end{equation}
the average valency in the graph.

Let us assume that we need to repeat the process of iterating the dynamics namely (26) or (27) for $s$
steps in order to achieve the goal(s) of the algorithm(s). These goals vary for the various problems of
interest.
For cluster identification we need the points representing clusters in the graph to physically converge
into definable separate regions.

For graph characterizations and comparison we need a fewer number of steps,
sufficient to make the distances $r_{ij}(t)$ vary considerably away from their original common value.

The total number of computations involved is $N=O(n^2 s)$ if $v$ is finite and $n$ independent or
$N=O(n^3 s)$ for the extreme case when $v\approx n$. For $N$ not to be polynomial in $n$ we need that
$s$ will grow faster than any power of $n$.

In principal one can envision many types of chaotic dynamical evolution where such large number of steps
is indeed required.

This is not the case for the first order equations considered here:\\
\begin{equation}
\dot{\vec{r_i}} =  \frac {\vec F_i}{\mu} = - \frac{\vec \nabla _{r_i}(U)}{\mu}, 
\end{equation}
where the system consistently moves, along the steepest descent, to a minimum of $U$, the potential
energy.

If we have a complicated ``energy landscape'' the system can be trapped in any one of the many local
minima, a feature which accounts for the difficulty of protein folding\cite{bio}, neural nets and spin glass
problems\cite{spin}. The need to keep the same deterministic evolution for $S$ and $S'$ representing $G$ and $G'$
in the first ``distortion'' algorithm, excludes in our case- the possibility of introducing some
stochastic noise to extricate the system from a local minimum.

Fortunately our problem does not allow for many minima. Thus let us fix the locations of all 
$\vec r_i$\ \ \ $i = 1, \cdots , n-1$ except $\vec r_n \equiv \vec r$. The velocity $\dot{\vec r}(t)$,
is dictated
by
\begin{equation}
U(\vec r) = \sum_{i = 1} C_{n_{i}}U_A (|\vec r - \vec r_i |) + (1- C_{n_{i}}) U_R (|\vec r - \vec r_i |)
\end{equation}
Assume we have some local equilibrium at $\vec{r}_0$. Locally, in the neighborhood of $\vec{r}_0$,  we can use the variables  
$\rho _i \equiv |\vec r - \vec r_i |$\ \ \ $i= 1, \cdots , n-1$, instead of 
$x_1 \cdots x_{n-1}$ the $n-1$
Cartesian coordinates of $\vec{r}$. The conditions for an extremum 
$\vec \nabla U(\vec r)\mid _{\vec r = \vec r_o}$ then require that\\
\begin{equation}
\frac{\partial}{\partial \rho _j}U_A (\rho _j)\mid _{\rho _j = \rho^{(o)} _j} = 0;\ \ \ or \ \ \  \frac{\partial}{\partial \rho _j}U_R (\rho _j)\mid _{\rho _j = \rho^{(o)} _j} = 0
\end{equation}
Thus for generic monotonic $U_A$, $U_R$, we have no extrema inside the region.\\
An absolute minimum obtained at the boundary.

\section{Applications of the method}

\begin{figure}[h]
\includegraphics{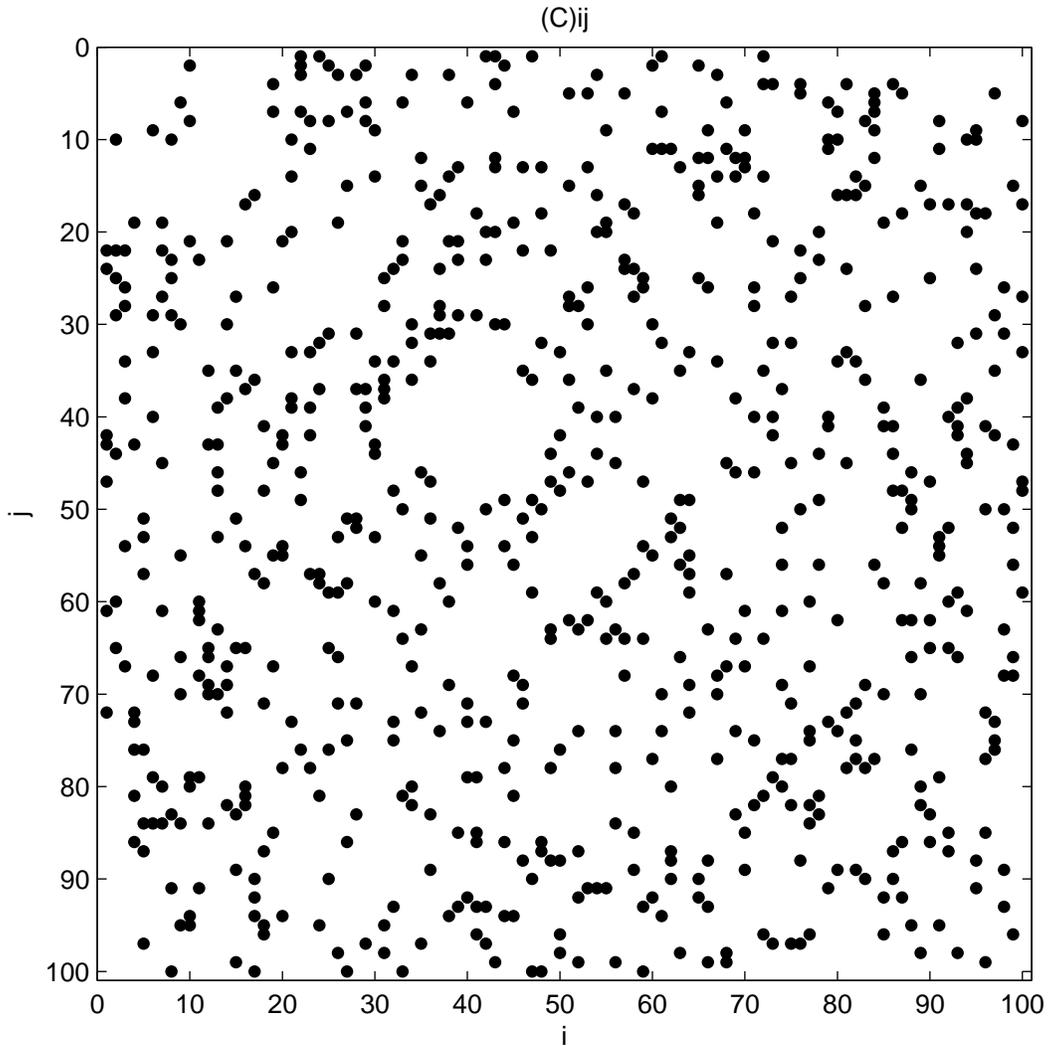}
\caption{Connectivity matrix for 100 vertices graph with 7 random connections for each vertex.}
\label{fig:c7}
\end{figure}

\begin{figure}[h]
\includegraphics{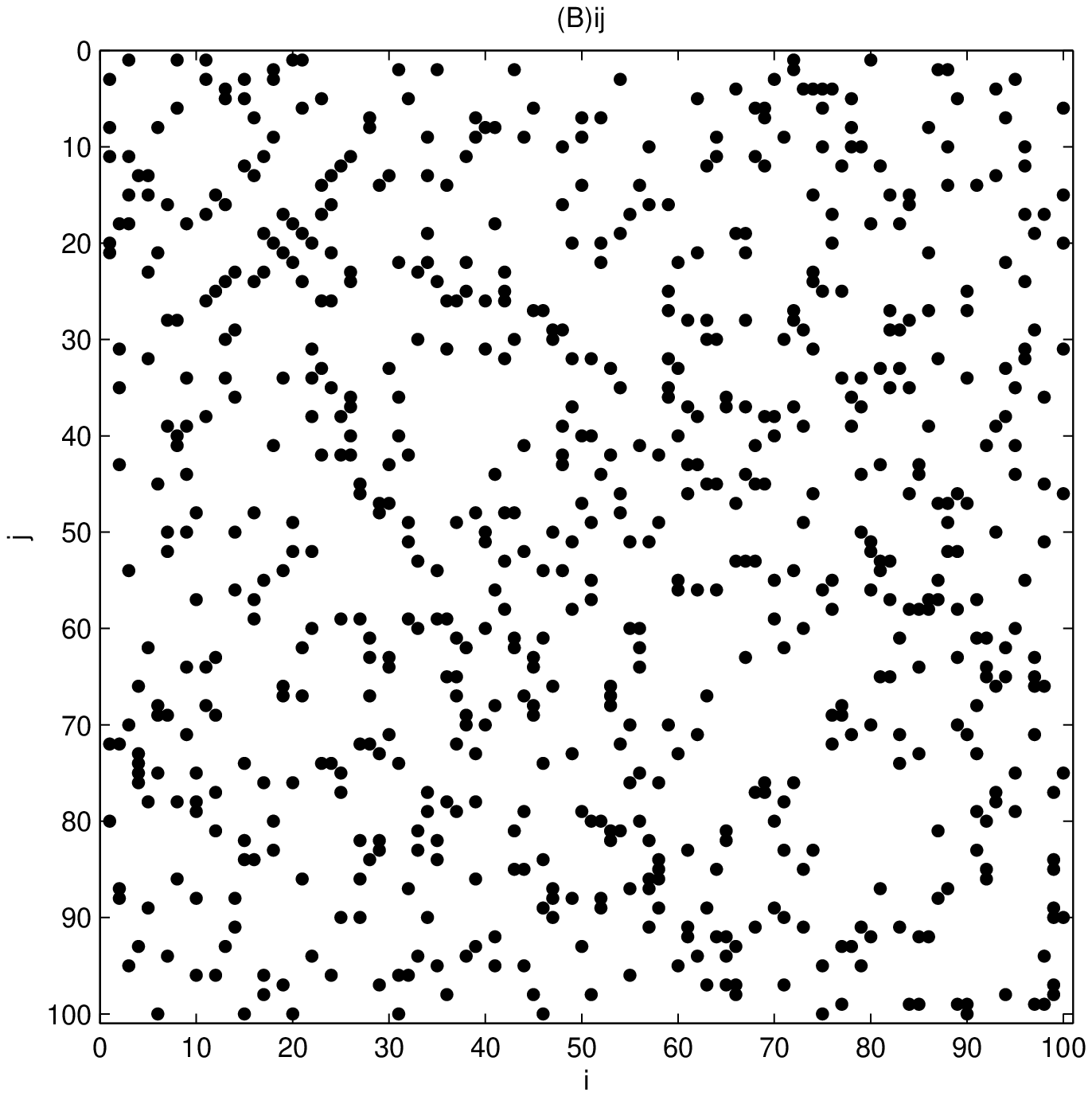}
\caption{Reshuffled connectivity matrix.}
\label{fig:b7}
\end{figure}

To demonstrate the power of our approach we considered a graph with 100 vertices each of which is randomly connected to seven others. The corresponding connectivity matrix $C$ is shown in Fig.(\ref{fig:c7})).  
Random reshuffling transforms the $C$  into the matrix $B$ of Fig.(\ref{fig:b7}) 
Next we  applied our algorithm using a combination of attractive and repulsive forces in $n-1=99$ dimensional space. The vertices of the 100-simplex were allowed to move under the influence of the forces on the 98-dimensional hyper-sphere in 99-dimensions. After a number of steps we analyzed the  distances between pairs the vertices of the two simplexes. We found perfect correspondence between the distance matrices. We also readily show the permutation matrix which maps one distance matrix on to the another. Applying the latter to the matrix $B$ reproduces exactly the original connectivity matrix $C$ (Fig.(\ref{fig:c7})).

\begin{acknowledgments}
\section{}
S.Nussinov would like to thank Zohar nussinov for a crucial comment 
 regarding the advantage of going to higher dimensions to overcome 
 frustrations and alleviate constraints. He would like to dedicate this work to sir Isac Wolfson who donated the chair in theoretical physics at Tel-Aviv University on the occasion of his 80th birthday.
\end{acknowledgments}


\begin{thebibliography}{}

\bibitem{gold}An analogous physical system was used by Farhi, Goldstone and Gutmann and Sipser arXiv quant-ph/0001106 (2000). Their idea was to create a eave function of $n$ spins satisfying a set of Boolean logic logic requirments via adiabatic changing of the Hamiltonian.
\bibitem {eigen} D. Cvetkovi\'{c}, P. Rowlinson and S. Simi\'{c}, ``Eigenspaces of graphs'' (Encyclopedia of mathematics and its applications, v. 66), Cambridge; New York : Cambridge University Press, 258p., 1997.
\bibitem {prob} A. R\'{e}nyi, ``Probability Theory'', North-Holland Publishing Company - Amsterdam - London, and American Elsevier Publishing Company, Inc. - New York, 1970.
\bibitem {cosp} C.D.Godsil, D.A. Holton and B.D. McKay, ``The spectrum of a graph'', Lecture Notes in Math. {\bf 622}, Springer-Verlag, Berlin, 1977,91-117.
\bibitem {coor} For a specific convenient coordinate choice for the $n$ vertices see Appendix of \cite{gjn}.
\bibitem {gjn} V. Gudkov, J.E. Johnson and S. Nussinov, arXiv: cond-mat/0209111 (2002).
\bibitem {bio} J.D. Bryngelson and P.G. Wolynes, J. Phys. Chem. {\bf 93}, p. 6902 (1989).
\bibitem {spin} M. M'ezard, G. Parisi and M.A. Virasoro, ``Spin glass theory and beyond'', World Scientific, Singapore, 1987.



\end{thebibliography}
\end{document}